\documentclass[journal]{IEEEtran}
\usepackage{amsmath,amsfonts}
\usepackage{algorithmic}
\usepackage{algorithm}
\usepackage{array}
\usepackage{textcomp}
\usepackage{stfloats}
\usepackage{url}
\usepackage{verbatim}
\usepackage{cite}
\usepackage{setspace}
\usepackage{amsthm}
\usepackage{amssymb,amsmath,latexsym}
\usepackage{color,cite,times}
\usepackage{epstopdf}
\usepackage{multirow}
\usepackage{array}
\usepackage{booktabs}
\usepackage{graphicx}
\usepackage{subcaption}

\newtheorem{remark}{\textbf{Remark}}
\allowdisplaybreaks[4]

\renewcommand{\aa}{\mathbf{a}}

\renewcommand{\ll}{\mathbf{l}}
\providecommand{\mm}{\mathbf{m}}

\providecommand{\pp}{\mathbf{p}}
\providecommand{\qq}{\mathbf{q}}

\providecommand{\vv}{\mathbf{v}}
\providecommand{\ww}{\mathbf{w}}

\ifCLASSINFOpdf
\else
\fi
\begin{document}
	
	\title{
Trajectory Design for UAV-Assisted Logistics Collection in Low-Altitude Economy}
	\author{
			Zhiyuan~Zhai, Yuan~Gao, Wei Ni, \IEEEmembership{Fellow, IEEE},
	Xiaojun~Yuan,~\IEEEmembership{Senior Member, IEEE}, 
	and Xin~Wang,~\IEEEmembership{Fellow, IEEE}
}

	\maketitle

\begin{abstract}
Low-altitude economy (LAE) is rapidly emerging as a key driver of innovation, encompassing economic activities taking place in airspace below  500 meters. Unmanned aerial vehicles (UAVs) provide valuable tools for logistics collection within LAE systems, offering the ability to navigate through complex environments, avoid obstacles, and improve operational efficiency. However, logistics collection tasks involve UAVs flying through complex three-dimensional (3D) environments while avoiding obstacles, where traditional UAV trajectory design methods—typically developed under free-space conditions without explicitly accounting for obstacles—are not applicable. This paper presents, we propose a novel algorithm that combines the Lin-Kernighan-Helsgaun (LKH) and Deep Deterministic Policy Gradient (DDPG) methods to minimize the total collection time. Specifically, the LKH algorithm determines the optimal order of item collection, while the DDPG algorithm designs the flight trajectory between collection points. Simulations demonstrate that the proposed LKH-DDPG algorithm significantly reduces collection time by approximately \textbf{49\%} compared to baseline approaches, thereby highlighting its effectiveness in optimizing UAV trajectories and enhancing operational efficiency for logistics collection tasks in the LAE paradigm.
\end{abstract}


\section{Introduction}
Low-altitude economy (LAE) refers to economic activities occurring in airspace below $500$ meters \cite{watson2020maximizing}, including logistics, transportation, agriculture, and surveillance \cite{wu2024realization}. LAE exploits underutilized low-altitude airspace to provide fast, flexible, and scalable services \cite{xiong2024evtol}. Not only does it enable new business models, such as aerial logistics and on-demand urban aerial mobility, but it also improves sustainability by alleviating the burden on ground infrastructure. Driven by modern technologies in autonomous flight, sensing, and artificial intelligence, LAE has emerged as an essential component of smart cities, offering solutions for urban delivery, transportation, disaster relief, and emergency response. Consequently, the design, analysis, and optimization of LAE systems, especially under realistic and obstacle-rich environments, have become a key research focus \cite{yang2025intelligent, tang2021systematic, hsu2022rl-uav_comm, wang2021learning_collision, gong2022bor_opt, sha2025rrt_sa_cost, xu2021dpm_pc, roghair2021vision_rl}.
\subsection{Motivation}
Among many applications of LAE, logistics collection is particularly promising, as it underpins time-sensitive tasks, such as parcel pick-up, medical supply delivery, and cargo transportation, in dense urban areas. Unmanned aerial vehicles (UAVs) \cite{fahlstrom2022introduction}, thanks to their excellent maneuverability, relatively low deployment cost, and autonomous operation capability, have been recognized broadly as a promising technology for such tasks \cite{zhai2025integrated,zhai2024uav}. UAVs are capable of navigating complex environments while avoiding obstacles, thereby reducing dependence on traditional ground-based transportation. Their ability to fly at low altitudes, combined with flexibility in accessing narrow streets, rooftops, and remote rural areas, makes UAVs highly effective in logistics collection. In addition, UAVs can substantially reduce delivery times, avoid ground traffic congestion, and reach otherwise inaccessible locations. These advantages make UAVs indispensable for future LAE-enabled logistics networks, improving both the efficiency and scalability of operations \cite{zhang2023threat_model, tu2023q_sarsa, shaoxiao2025airspace_comm,9169676}.

\subsection{Challenges}
For UAV-assisted logistics collection to be practically viable, careful trajectory design is indispensable. Existing works on UAV trajectory optimization typically assume free-space operations, where obstacles are overlooked, simplified, or modeled only implicitly \cite{wu2018joint,zhai2022energy,7888557,10818523,sun2021joint,xiong2022three}. While these assumptions allow analytical simplifications, they are far from realistic in low-altitude urban environments filled with buildings, vegetation, and other infrastructures. Some studies, e.g., \cite{banerjee2021uav_risk, nzhang2021autonomous_collision,10480601}, have extended trajectory designs into three-dimensional (3D) spaces, but most have focused on altitude control or communication-related objectives without explicitly accounting for physical obstacle avoidance. In other words, they have implicitly assumed that collision risks are absent,  significantly limiting their applicability to real-world LAE scenarios.

In practice, UAVs performing logistics collection often encounter diverse and dense obstacles at low altitudes. The UAVs must not only determine efficient paths to visit all collection points but also dynamically adapt their trajectories to avoid collisions, maintain safety, and ensure task completion. This makes the trajectory design fundamentally more complex than free-space operations, as it involves a highly non-convex feasible region coupled with discrete–continuous decision variables. To this end, there is an urgent need for new approaches that explicitly consider obstacles, capture the non-convexity of the feasible domains, and enable UAVs to flexibly adjust both altitude and horizontal positions in real time \cite{ma2023flightcorridor, chan2025nearoptimal, chao2025airspace, hu2023droneswarm}.

\subsection{Related Works}
A large body of existing UAV trajectory design research has been carried out under free-space assumptions, where the propagation environment is modeled without physical obstacles and light-of-sight (LoS) links are always available.  
For instance, Wu \emph{et al.} \cite{wu2018joint} investigated joint trajectory and resource allocation for multi-UAV downlink communication, formulating a non-convex problem under the assumption of unobstructed propagation. Zeng and Zhang \cite{7888557} studied energy-efficient UAV communication by jointly optimizing the UAV's trajectory and transmit power to minimize its propulsion energy consumption, exploiting the analytical tractability of free-space channels.  
Lyu \emph{et al.} \cite{lyu2018uav_jsac} designed UAV-mounted base station placement and flight paths to maximize user coverage in hotspot areas, relying on an ideal LoS model.  
Zhang \emph{et al.} \cite{zhang2019sec_twc} developed secure UAV communication schemes via joint trajectory and power control to enhance secrecy rates, again under obstacle-free assumptions. 
 
In addition to communication-centric studies, UAV trajectory design has also been applied to computation-oriented tasks. For example, Zhai \emph{et al.} \cite{zhai2022energy} proposed an energy-efficient UAV-mounted reconfigurable intelligent surface (RIS)-assisted mobile edge computing framework, where the UAVs' trajectory was optimized to balance computation offloading and energy efficiency.  
Sun et al. \cite{sun2021joint} studied UAV-assisted edge computing, jointly optimizing the UAV’s trajectory and computation offloading to minimize energy consumption.
Although these works differ in objectives—ranging from energy efficiency to coverage, secrecy, and offloading performance—they all share the simplified assumption of free-space operations without explicit obstacle modeling.

While the above free-space formulations simplify optimization and enable closed-form analysis, they are inadequate for realistic LAE environments.  
In dense urban airspaces, UAVs must navigate through complex 3D obstacle distributions, including buildings, vegetation, and other infrastructures.  
These obstacles induce highly non-convex feasible regions, compromising the validity of straight-line flight segments assumed in free-space models and significantly complicating trajectory planning.  
Algorithms developed under idealized conditions can yield infeasible or unsafe paths.  
To this end, our work departs from the traditional free-space paradigm by explicitly incorporating obstacle-avoidance constraints into both problem formulation and solution design. We focus on the UAV-enabled logistics collection scenario, where the UAVs must determine not only the optimal order of item collection but also dynamically feasible, collision-free trajectories in obstacle-rich 3D urban environments.

\subsection{Contributions}
In this paper, we consider a UAV-enabled logistics collection system, where a UAV is dispatched from a central depot, navigates around obstacles, and collects items distributed in the environment, before returning to the depot. We minimize the overall mission completion time while satisfying kinematic and obstacle-avoidance constraints. Unlike traditional trajectory design approaches typically under idealized assumptions, this problem is highly non-convex due to three key challenges: (i) The objective function is directly coupled with the trajectory duration; (ii) the feasible set is defined by strict obstacle-avoidance constraints that cannot be convexified by standard methods; and (iii) the collection constraints involve non-linear intersection operations. These difficulties render classical convex optimization and control-theoretic methods, e.g., model predictive control, inadequate for efficient real-time trajectory design.

To address these challenges, we propose a novel algorithm that integrates the Lin-Kernighan-Helsgaun (LKH) traveling salesman problem (TSP) solver \cite{helsgaun2015solving} with the Deep Deterministic Policy Gradient (DDPG) algorithm \cite{lillicrap2015continuous}. The LKH component determines an efficient visiting order of items, while the DDPG component adaptively learns continuous flight trajectories between successive items in the presence of obstacles.  This hybrid approach, termed {LKH-DDPG}, exploits the complementary strengths of LKH (for combinatorial sequence optimization) and DDPG (for continuous trajectory optimization), making it particularly suitable for obstacle-rich LAE scenarios.

The  contributions of this paper are summarized as follows:
\begin{itemize}
	\item We investigate the UAV-assisted logistics collection problem in the emerging LAE paradigm, explicitly accounting for UAV kinematic constraints, strict obstacle avoidance, and item collection requirements. 
	\item  We decompose this highly non-convex optimization problem into two sub-problems: (i) discrete visiting order optimization solved using the LKH algorithm, and (ii) continuous obstacle-aware trajectory design solved using DDPG.
	\item  We integrate global visiting order optimization with local trajectory learning into a unified framework, leveraging the complementary strengths of combinatorial optimization and reinforcement learning. This framework explicitly addresses the hybrid discrete–continuous nature of the problem and ensures adaptability in complex 3D environments.
	\item  Simulations demonstrate that the proposed LKH-DDPG method enables a UAV to collect items smoothly and collision-free while substantially shortening mission duration. For example, with $K=8$ items, LKH-DDPG achieves a mission time of $97.0$~s, shortening time by approximately $45.7\%$ compared to DDPG without LKH ($178.8$~s) and $49.4\%$ compared to model predictive control (MPC) ($191.8$~s).
\end{itemize}

The rest of this paper is organized as follows. Section~II describes the system model, including UAV trajectory representation, kinematic constraints, item collection, and obstacle modeling. Section~III formulates the mission time minimization problem. Section~IV presents the proposed LKH–DDPG framework, where LKH optimizes the collection sequence and DDPG designs obstacle-aware trajectories. Section~V presents numerical results, comparing the proposed method with baselines.  Section~VI concludes the paper.

\textit{Notation:} 
Italic letters denote scalar variables;  bold lowercase and uppercase letters represent vectors and matrices, respectively; $(\cdot)^\top$ denotes transpose, $\|\cdot\|$ stands for the Euclidean norm, and $|\cdot|$ denotes the absolute value or set cardinality, depending on the context;  $\dot{(\cdot)}$ and $\ddot{(\cdot)}$ denote the first and second derivatives, respectively.

\section{System Model}
\begin{figure}[htbp]
	\centering
	\includegraphics[width=0.4\textwidth]{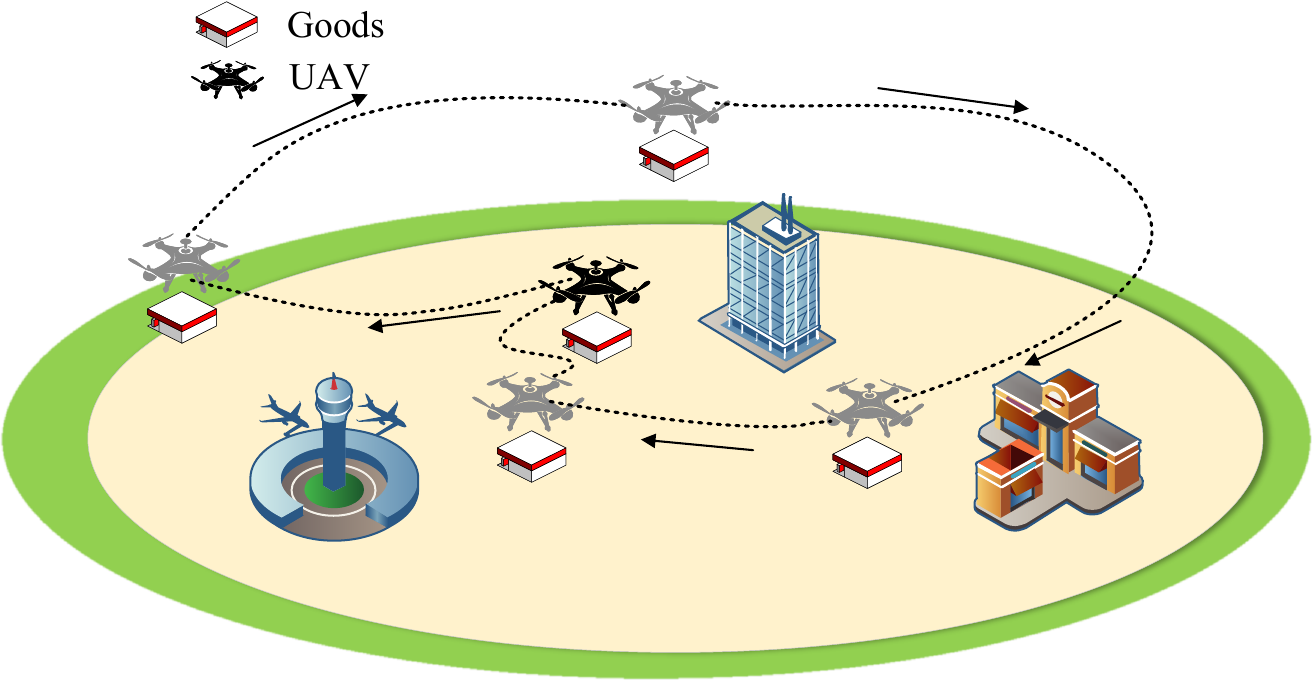}
	\caption{An example of the UAV-assisted LAE system. 
		A UAV is dispatched from the depot, visits multiple item locations distributed in a 3D urban environment with obstacles (e.g., buildings), collects items, and finally returns to the depot within the specified delay $T$.}	
	\label{fig_system_model}
\end{figure}

We investigate a logistics collection task in the context of  LAE, focusing on environments with diverse and complex spatial characteristics. As illustrated in Fig.~\ref{fig_system_model},  a UAV executes a task of collecting   $K$  items from $K$ users distributed across a range of different environments. The environments may include urban areas with dense buildings, suburbs with varying land use, or mountainous terrains with significant elevation changes. The UAV needs to navigate through complex environments to complete item collection.

The entire system is modeled in a three-dimensional (3D) Cartesian coordinate system to capture the spatial position and variations of the environment. Each user $k$ is located at a fixed position  $\mathbf{w}_k = [x_k, y_k, z_k]^\top$, where $x_k$, $y_k$, and $z_k$ represent the user's coordinates along the respective axes. During the UAV's flight,  its trajectory at time instant $t$ is given by $\mathbf{q}(t), 0 \leq t \leq T$, where  $\mathbf{q}(t) = [x(t), y(t), z(t)]^\top$ denotes the UAV's position at time  $t$. 

In the considered system, the UAV operates based on a central workstation, which serves as both the launch and docking points. This workstation provides a facility for the UAV to refuel, recharge, or offload collected items. The UAV's trajectory must adhere to the following constraint:
\begin{equation}\label{c1}
	\mathbf{q}(0) = \mathbf{q}(T),
\end{equation}
which implies that the UAV needs to return to the workstation by the end of the period \(T\).  

In practical applications, the UAV's trajectory must also adhere to constraints on the maximum speed\footnote{The minimum speed constraints are not considered for rotary-wing UAVs, which are capable of hovering at fixed positions, where a zero-speed state is feasible. For fixed-wing UAVs requiring forward motion to maintain flight, an additional constraint \(\|\dot{\mathbf{q}}_m(t)\| \geq V_\text{min} > 0, \, 0 \leq t \leq T, \, \forall m\) must be enforced. This can be accommodated by the proposed algorithm with minor modifications.} and acceleration, formulated as
\begin{equation}\label{c2}
	\|\dot{\mathbf{q}}(t)\| \leq v_\text{max}, ~	\|\ddot{\mathbf{q}}(t)\| \leq a_\text{max}\quad 0 \leq t \leq T, 
\end{equation}
where  \(v_\text{max}\) and $a_\text{max}$ are the maximum allowable UAV speed and acceleration, respectively.
To facilitate analysis, the duration \(T\) is divided into \(N\) uniform time intervals indexed by \(n = 1, \ldots, N\), with each interval  length of \(\delta = \frac{T}{N}\). This interval is assumed to be short enough that the UAV's position is approximately unchanged within a slot, even at the maximum speed of the UAV \(v_\text{max}\). Consequently, the UAV's trajectory is represented by a set of \(N\) 3D points, $\qq[n]=\left[x[n],y[n],z[n]\right], n=1,\cdots,N$. Constraints   \eqref{c1} and \eqref{c2} can be rewritten as
\begin{align}
	\mathbf{v}[n] &= \frac{\mathbf{q}[n+1] - \mathbf{q}[n]}{\delta}, \quad \|\mathbf{v}[n]\| \leq v_\text{max}, \, \forall n; \label{2}\\
	\mathbf{a}[n] &= \frac{\mathbf{v}[n+1] - \mathbf{v}[n]}{\delta}, \quad \|\mathbf{a}[n]\| \leq a_\text{max}, \, \forall n;\label{3}\\
	\qq[1]&=\qq[N].\label{4}
\end{align}

During the logistics collection process, the UAV navigates through obstacles. It can only successfully collect item $i$ at time slot $n$ when the following conditions are satisfied:
\begin{align}\label{5}
	\mathbf{v}[n]=0, ~~\|\qq[n]-\ww_i\|\leq \epsilon,
\end{align}
where a speed of zero indicates that items can only be collected when the UAV is hovering;  the condition \(\|\qq[n]-\ww_i\|\leq \epsilon\) signifies that the UAV must be reasonably close to item $i$. Here, $\epsilon$ defines the collecting threshold. 

Apart from the flight and item collection mechanisms,  in this system, a key consideration is that the UAV must avoid obstacles in the environment during its flight. 
We model the $i$-th obstacle in the environment as a rectangular cuboid, defined by two vertices on its diagonal,  \( \mathbf{p}_{\text{min}}^i=[x_{\text{min}}^i, y_{\text{min}}^i, z_{\text{min}}^i] \) and \( \mathbf{p}_{\text{max}}^i=[x_{\text{max}}^i, y_{\text{max}}^i, z_{\text{max}}^i] \). Here, \( \mathbf{p}_{\text{min}}^i \) represents the vertex  closer to the origin of the considered 3D coordinate system, while \( \mathbf{p}_{\text{max}}^i \) represents the vertex  farther from the origin. 

To prevent the UAV from colliding with these obstacles, its trajectory is constrained such that it does not enter the interior of the rectangular cuboids. For  obstacle $i$, this condition can be expressed as
\begin{align}
&	x[n] < x_{\text{min}}^i \,~ \text{or} \, x[n] > x_{\text{max}}^i, \forall i;\label{6}\\
&	y[n] < y_{\text{min}}^i \,~ \text{or} \, y[n] > y_{\text{max}}^i, \forall i; \label{7}\\
&	z[n] < z_{\text{min}}^i \,~ \text{or} \, z[n] > z_{\text{max}}^i,\forall i.\label{8}
\end{align}

These constraints guarantee that the UAV always maintains
a safe distance from obstacles during its flight. Note that for
each obstacle, the conditions in \eqref{6}-\eqref{8} are connected by a
logical {OR}, meaning that it is sufficient for the UAV to
stay outside the obstacle along at least one dimension. As a
result, the UAV’s trajectory must be carefully designed within
the constrained 3D space, ensuring both obstacle avoidance
and task feasibility/completion.

\section{Problem Formulation}
In this system,  the UAV must be dispatched from the workstation, approach each item to retrieve it, and then return to the workstation with all collected $K$ items. We  minimize the time required for the UAV to complete the task by formulating the following task completion time minimization problem:
\begin{subequations}\label{problem1}
	\begin{align}
		\min_{\mathbf{Q}}& \quad T\label{10a}\\
		\text{s.t.}& \quad \eqref{2},\eqref{3},\eqref{4},\eqref{6},\eqref{7},\eqref{8},\\
			&\quad \sum_{n=1}^{N}\sum_{i=1}^{K} (\vv[n]=0\cap\|\qq[n]-\ww_i\|\leq \epsilon)=K,\label{9c}
\end{align}
\end{subequations}
where  $\cap$ is the intersection operator. The condition $(\vv[n]=0 \cap \qq[n]=\ww_i)$ takes $1$ if and only if the UAV is hovering ($\vv[n]=0$) and its position coincides with the location of item $i$ ($\qq[n]=\ww_i$) at time slot $n$; otherwise, the condition takes $0$. Constraint \eqref{9c} ensures that all $K$ items are collected exactly once during the mission.

Problem \eqref{problem1} is a challenging non-convex optimization problem, with its complexity arising primarily from the following aspects. On the one hand, the objective \eqref{10a}  depends closely on the number of time slots  $N$. Since $N$  depends on the discretized trajectory length, the optimization directly involves the trajectory duration as a variable. This coupling between the objective function and the optimization variables makes the problem strongly non-convex and particularly challenging.

On the other hand,  unlike typical UAV trajectory design problems \cite{wu2018joint,zhai2022energy,7888557,10818523,sun2021joint,xiong2022three}, the UAV in our system must fly between obstacles with a trajectory satisfying constraints \eqref{6}, \eqref{7}, and \eqref{8}. These constraints are non-convex \cite[Sec.~2.1.4]{boyd2004convex}. Methods, such as successive convex approximation (SCA) \cite{liu2019stochastic},  can convexify non-convex constraints. However, in our system, the feasible trajectories can intersect with the surfaces of the cuboid obstacles, forming a complex space that cannot be readily characterized by a single surrogate function. Consequently, these methods cannot convexify \eqref{6}--\eqref{8},  further increasing the difficulty of solving the problem.

Another challenge in solving problem \eqref{problem1} lies in constraint \eqref{9c}. To determine whether the UAV successfully collects an item at a given location, constraint \eqref{9c} involves intersection operations, which are inherently non-convex. 

In the subsequent section, we develop a novel algorithm tailored to problem \eqref{problem1} for optimizing the UAV's trajectory.

\section{Proposed LKH-DDPG Algorithm}
In this section, we delineate the proposed LKH-DDPG algorithm to design the UAV's trajectory for item collection in the LAE scenarios. As analyzed in Section~III, the formulated problem is highly
non-convex due to the coupled trajectory duration, strict
obstacle-avoidance constraints, and non-linear item collection
conditions. These difficulties make conventional optimization-
based approaches inadequate for efficient trajectory design.
To this end, we turn to deep learning methods. Reinforcement
learning (RL), in particular, has been widely recognized as an
effective tool for UAV trajectory optimization in complex
environments \cite{marini2022reinforcement,ning2023multi,gong2023bayesian}, since it can directly learn control policies from
interaction with the environment.

However, applying RL alone to the studied logistics
collection problem faces a key challenge. While RL algorithms
such as DDPG can effectively design obstacle-aware
continuous trajectories, they struggle to simultaneously
determine the optimal visiting order of items. Without a given
collection sequence, the learning process must handle both
the discrete combinatorial optimization of visiting order and
the continuous trajectory control, which often leads to poor
convergence or infeasible solutions \cite{roghair2021vision,yan2020towards}. To overcome this issue,
we integrate the LKH algorithm,
which efficiently determines a near-optimal visiting sequence,
with DDPG, which optimizes the continuous trajectories
between collection points. This hybrid design enables the UAV
to achieve both global sequence optimization and local
obstacle-aware trajectory learning.

\subsection{Collecting Sequence Design}
To solve problem \eqref{problem1}, we first need to determine the order in which the UAV visits and collects items. This task can be naturally modeled as a classical  TSP \cite{hoffman2013traveling}, where the UAV  must visit each location exactly once before returning to the workstation. The TSP formulation aims to find the order of visits that minimizes the total travel distance. In this step, we approximate the distance between any two items by a straight-line segment without considering the obstacles. 
The TSP   is formulated as
\begin{subequations}\label{10pro}
	\begin{align}
			&\min_{u_i, \forall i,x_{ij},\forall i,j} \quad \sum_{i=1}^{K} \sum_{\substack{j=1 \\ j \neq i}}^{K} d_{ij} x_{ij}\\
			\text{s.t.} \quad 
			&\sum_{j=1, j \neq i}^{K} x_{ij} = 1, \quad \forall i \in [K], \label{111}\\
			&\sum_{i=1, i \neq j}^{K} x_{ij} = 1, \quad \forall j \in [K], \label{222}\\
			&u_i - u_j + K x_{ij} \leq K - 1, \quad 2 \leq i \neq j \leq K, \label{13a}\\
			&x_{ij} \in \{0,1\}, \quad \forall i,j \in \mathcal{V}, \; i \neq j,\\
			&2 \leq u_i \leq K, \quad \forall i=2,\dots,K.
		\end{align}
\end{subequations}
where, $x_{ij},\forall i,j$ is a binary decision variable: $x_{ij}=1$, if the UAV directly travels from item $i$ to item $j$, and $x_{ij}=0$, otherwise;
 $d_{ij}$ denotes the Euclidean distance between items $i$ and $j$;
 $u_i,\forall i$ represents the order of visiting item $i$ in the collection tour 
(Without loss of generality, we set $u_1 = 1$ to represent the starting depot, i.e., workstation, while the remaining variables $u_i \in \{2,\dots,K\}$ enforce a valid order of visits); $\mathcal{V} = \{1,2,\dots,K\}$ denotes the set of all nodes (items) in the problem. 
Constraints \eqref{111} and \eqref{222} enforce that each item has exactly one incoming path and one outgoing path, thereby ensuring a complete tour. 
Constraint \eqref{13a}  eliminates potential sub-tours \cite{sawik2016note}, so that all items are connected within a single feasible route. 

To efficiently solve the TSP, we adopt the LKH algorithm \cite{helsgaun2015solving}, which is a powerful heuristic solver for large-scale TSP instances. LKH is based on the Lin–Kernighan local search and improves candidate tours through variable-depth $k$-opt edge exchanges. By dynamically adjusting $k$ and exploiting efficient candidate sets, LKH can effectively escape local minima and produce a near-optimal order of visits, i.e., $x_{ij},\forall i,j$. This makes LKH particularly suitable for determining the UAV’s item-collecting order in our problem setting.

\subsection{Flight Trajectory Design}
Given the UAV's item-collecting order determined by LKH, we design the flight trajectory between the items visited successively. 
Unlike conventional free-space trajectory optimization,  problem  \eqref{problem1} must be solved in a highly non-convex feasible region, where the UAV’s motion is constrained by strict obstacle-avoidance requirements even under the given item collecting order. 
This makes direct optimization extremely challenging. 
To circumvent this impasse, we model the trajectory design as a sequential decision-making problem. In each step, the UAV must determine its control action by considering its current position, velocity, the surrounding obstacles, and the relative location of the next item. 
Such a formulation naturally motivates the use of reinforcement learning that can adaptively handle continuous state and action spaces in complex non-convex environments.

DDPG, developed by DeepMind  \cite{lillicrap2015continuous}, is a suitable approach for solving the trajectory design problem. DDPG is an off-policy deep reinforcement learning (DRL) algorithm that excels in environments with continuous action spaces, making it well-suited for UAV trajectory optimization.
The DDPG algorithm builds upon the actor-critic framework, combining the strengths of both the policy gradient and the value function methods. The policy function, denoted as \( \pi \), is typically referred to as the actor, while the value function, denoted as \( q\), is known as the critic. The actor generates actions based on the current state of the environment, selecting them from a continuous action space. The action \( A = \pi(S;\boldsymbol{\theta})\) is the UAV's acceleration. The critic outputs \( q(S, A;\ww) \), which evaluates the actions taken by the actor in relation to the current state of the environment. 

To apply DDPG, we define the state, action, and reward as follows.

\begin{enumerate}
	\item {\emph{State:}} The state of the considered system is given by $S=\{\qq,\vv,\ll,\mm,\pp\}$, where \( \ll \) represents the relative position vector from the UAV to each obstacle, \( \mm \) denotes the relative position vector from the UAV to each item, and \( \pp \) represents the collection status of the items. 
	\item {\emph{Action:}} The action $A$ is the UAV's acceleration. $A=\{\aa[n]\}$.
	\item {\emph{Reward:}} The reward \(R\) consists of six components: \(r_1\), the reward based on the distance between the UAV and an item;
	\(r_2\),  the reward for successfully collecting an item;
	\(r_3\),  the reward based on the distance between the UAV and the depot after collecting all items;
	\(r_4\),  the reward of returning to the depot after collecting all  items;
	\(r_5\),  the penalty for colliding with the obstacles; and
	\(r_6\),  the penalty for flying beyond the boundary.	
\end{enumerate}

\begin{algorithm}[h]
	\caption{DDPG  for  Trajectory Design}
	\label{algo1}
	\begin{algorithmic}[1]
		\STATE \textbf{Parameter:} Learning rate $\alpha^{(\ww)}$ and $\alpha^{(\theta)}$, discount factor $\gamma$, the number of episodes $M$, the number of steps per episode $P$, the number  of operations  executed when accumulating experience $U_1$,  the number of  operations executed when updating the parameter $U_2$, target network learning rate $\alpha_{\text{target}}$;
		\STATE \textbf{Initialization:} Initialize \(\boldsymbol{\theta}\) and \(\ww\), with \(\boldsymbol{\theta}_{\text{target}} \leftarrow \boldsymbol{\theta}\) and \(\ww_{\text{target}} \leftarrow \ww\), and   experience replay buffer $\mathcal{D} \leftarrow \varnothing $;
		\FOR {episode = $1$, $\dots$, $M$}
		\STATE Choose the  initial state  $S$;
		\FOR {$p$ = 1, $\dots$, $P$}
		\FOR {$u_1$ = 1, $\dots$, $U_1$}
		\STATE Apply perturbation to \( \pi(S;\boldsymbol{\theta}) \) (e.g., a  Gaussian process) to determine the action \( A \);
		\STATE Execute action \( A \), observe reward \( R \), new state \( S' \), and episode termination indicator \( D' \);
		\STATE Store the experience \(\{S, A, R, S', D'\}\) in the experience replay buffer \( \mathcal{D} \);

		\STATE $S\leftarrow S'$;
		\ENDFOR
		\FOR {$u_2$ = 1, $\dots$, $U_2$}
		\STATE Sample a batch of experiences \( \mathcal{B} \) from the storage buffer \( \mathcal{D} \), where each experience is in the form of \( \{S, A, R, S', D'\} \);
		\STATE Calculate temporal difference  return 
		\begin{align*}
			U \leftarrow R + \gamma q\!\left(S', \pi(S',\boldsymbol{\theta}_{\text{target}}); \ww_{\text{target}}\right)(1-D');
		\end{align*}
		
	\STATE Update 
	\begin{align*}
		\ww \leftarrow \ww &+ \alpha^{(\ww)} \frac{1}{|\mathcal{B}|}
		\sum_{\{S, A, R, S', D'\}\in \mathcal{B}}  \notag \\[-2pt]
		&\quad \times [U - q(S,A;\ww)] \nabla q(S,A;\ww);
	\end{align*}
		\begin{align*}
			\boldsymbol{\theta} \leftarrow \boldsymbol{\theta} 
			&+ \alpha^{(\theta)} \frac{1}{|\mathcal{B}|} 
			\sum_{\{S, A, R, S', D'\}\in \mathcal{B}}  
			\nabla \pi(S;\boldsymbol{\theta}) \notag \\[-2pt]
			&\quad \times 
			\bigl[\nabla_a q(S,a;\boldsymbol{w})\bigr]_{a=\pi(S;\boldsymbol{\theta})};
		\end{align*}
		
		\STATE Update 
		\begin{align*}
			\ww_{\text{target}} &\leftarrow (1-\alpha_{\text{target}})\ww_{\text{target}} + \alpha_{\text{target}}\ww; \\[2pt]
			\boldsymbol{\theta}_{\text{target}} &\leftarrow (1-\alpha_{\text{target}})\boldsymbol{\theta}_{\text{target}} + \alpha_{\text{target}}\boldsymbol{\theta};
		\end{align*}
		
		\ENDFOR
		\ENDFOR
		\ENDFOR
		\STATE \textbf{Output:} The optimal policy estimation $\pi(\boldsymbol{\theta})$.
	\end{algorithmic}
\end{algorithm}

We proceed to describe the DDPG algorithm tailored for the UAV flight trajectory design in complex obstacle-rich environments. The overall procedure is summarized in \textbf{Algorithm~1}. The core idea is to train an \emph{actor–critic} architecture with continuous state and action spaces. The actor network $\pi(S;\theta)$ outputs the control actions (i.e., the UAVs' accelerations) given the observed system state $S$, while the critic network $q(S,A;\omega)$ evaluates the quality of the chosen actions by estimating the corresponding Q-values. This design allows the algorithm to directly learn continuous-valued control strategies, which are essential for UAV trajectory optimization.  

The training process alternates between \emph{experience accumulation} and \emph{experience utilization}. In the experience accumulation phase (Steps 6–-10), the UAV interacts with the environment, executes actions perturbed by exploration noise (e.g., Gaussian processes), and collects transitions $\{S,A,R,S',D'\}$ that record the state, action, reward, next state, and termination flag. These transitions are stored in a replay buffer to break the temporal correlation among samples.  

In the experience utilization phase (Steps 13–-17), mini-batches of past experiences are sampled from the replay buffer for training. Step 14 computes the temporal-difference (TD) target using the target networks, which provide a stable reference for value learning. In Step 15, the critic parameters $\omega$ are updated by minimizing the mean-squared error (MSE) between the predicted Q-values and the TD target. This update ensures that the critic accurately evaluates the long-term return of each action. In Step 15, the actor parameters $\theta$ are updated according to the deterministic policy gradient theorem~\cite{lillicrap2015continuous}, which leverages the critic’s gradient information to improve the policy toward higher-value actions. Finally, in Step 16, the target actor and target critic networks are softly updated as weighted averages of the current networks. This technique stabilizes training by preventing drastic oscillations in the target values.  

By combining these mechanisms—the actor–critic structure, experience replay, and target networks—DDPG achieves stable and efficient policy learning in continuous action spaces. This makes it particularly suitable for UAV trajectory design, where the UAV must generate smooth acceleration commands in real time while navigating non-convex feasible regions and avoiding obstacles.

\begin{figure}[h]
	\centering                    
	\includegraphics[width=1\linewidth]{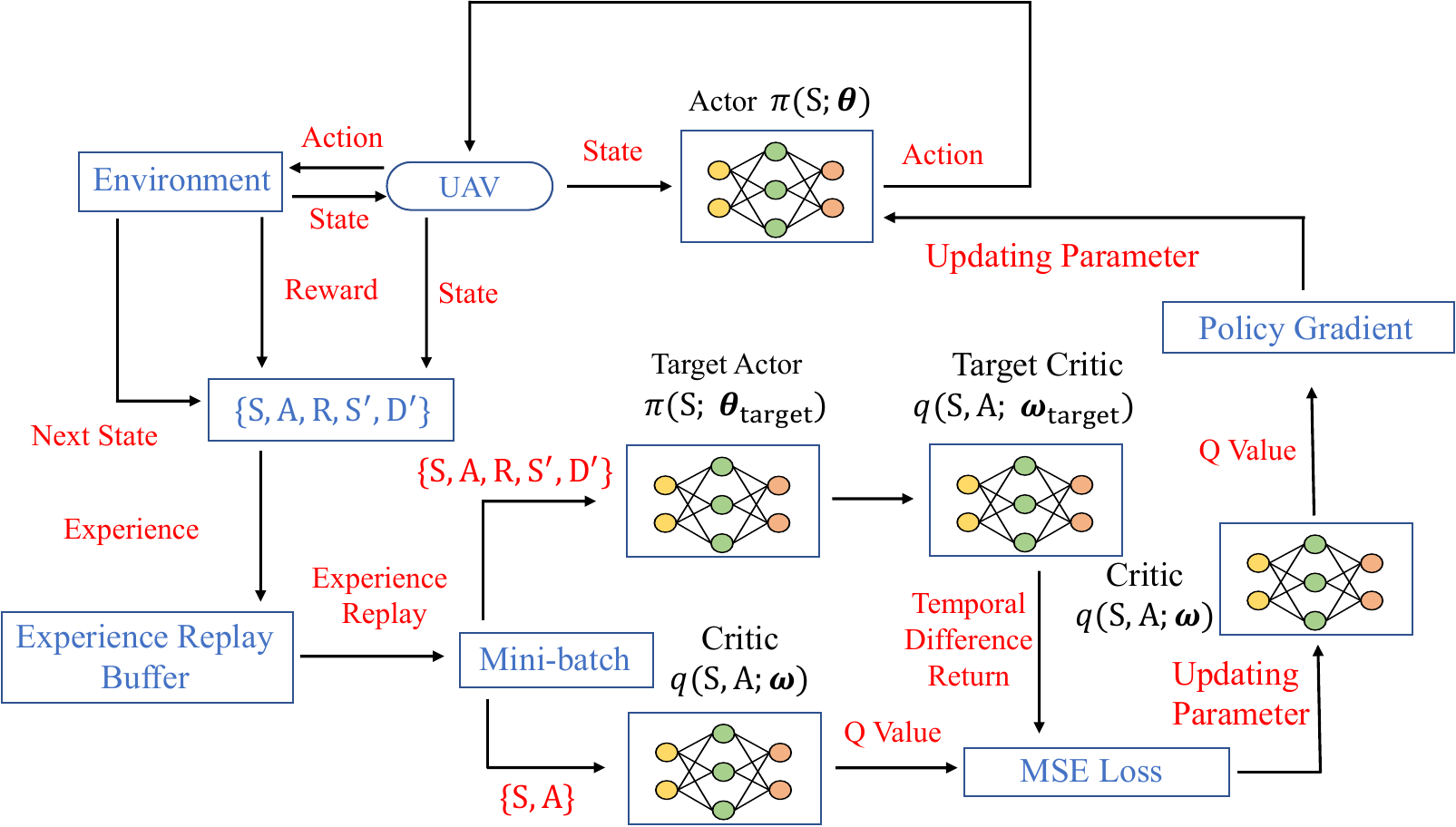}   
\caption{Workflow of the DDPG algorithm for UAV trajectory design. 
	The UAV interacts with the environment to generate experience tuples, which are stored in the replay buffer. 
	Mini-batches are sampled from the replay buffer to update the critic via TD learning. The actor is updated by policy gradients with target networks, ensuring stable training.}

	\label{DDPG_figure}
\end{figure}

Fig.~\ref{DDPG_figure} illustrates the integration of the DDPG framework into UAV trajectory design. 
In each step, the UAV observes the current state (e.g., relative positions of obstacles and collection targets) and selects a continuous acceleration action via the actor network. 
The resulting state–action–reward tuples are stored in a replay buffer for training, enabling the critic to evaluate action values and guide policy updates. 
Meanwhile, target actor and target critic networks are maintained to provide stable reference values during learning, thereby preventing oscillations in training. 
By iteratively refining the policy through interaction and evaluation, the UAV learns smooth, obstacle-aware flight paths tailored to the logistics collection task.

\subsection{Overall Algorithm}
We summarize the proposed UAV flight trajectory in \textbf{Algorithm \ref{algo2}}. 
The algorithm first initializes the UAV’s control parameters, including its speed, acceleration, item locations, and obstacle boundaries. 
It then employs the LKH algorithm to determine an efficient visiting sequence of items. 
Given this sequence, the DDPG-based trajectory planner (\textbf{Algorithm~1}) is applied to design the obstacle-aware flight paths between consecutive items and finally back to the depot. 
The output is a complete trajectory that jointly incorporates both global order optimization and local trajectory learning.

\begin{remark}
	The proposed LKH-DDPG algorithm is specially designed to address the unique difficulties of the studied UAV trajectory design problem in the presence of obstacles. Unlike conventional trajectory optimization methods, which typically assume free-space conditions \cite{wu2018joint,zhai2022energy,7888557,10818523,sun2021joint,xiong2022three}, our problem involves explicit obstacle avoidance in complex 3D spaces and an objective function directly coupled with the trajectory duration, making the problem strongly non-convex. Pure TSP solvers can determine only the order of visits but cannot adapt the continuous flight trajectory, while reinforcement learning alone may suffer from inefficiency due to the combinatorial nature of the order of visits. In the proposed LKH-DDPG framework, the LKH algorithm first generates an efficient visiting order of collection points.
	Given this order, the DDPG algorithm independently learns obstacle-aware continuous trajectories between consecutive points, ensuring safety and smoothness in complex environments.  
	This sequential integration allows each component to focus on its own strengths—LKH for combinatorial route optimization and DDPG for continuous control—resulting in a practical and effective solution for the studied logistics collection task.
\end{remark}

\begin{remark}
	The proposed LKH–DDPG algorithm operates in  {off-line}, i.e., it plans the UAV's trajectory beforehand. 
	Specifically, the LKH-DDPG algorithm assumes that the locations of all obstacles and collection items are known. 
	It first acquires this global information to construct the problem instance, solves the TSP using the LKH algorithm to obtain the visiting order, and then trains the DDPG agent to generate obstacle-aware trajectories for the UAV. 
	Due to the strongly non-convex nature of the problem caused by the coupling between mission time and trajectory variables, the strictly non-convex obstacle-avoidance constraints, and the non-linear intersection conditions for item collection, classical convex optimization, e.g., SCA,  cannot directly yield feasible solutions. 
	The proposed LKH-DDPG framework integrates global combinatorial optimization and local continuous policy learning,  making it particularly suitable for planning trajectories for logistics operations in complex 3D environments.
\end{remark}

\begin{algorithm}[h]
	\caption{LKH-DDPG}
	\label{algo2}
	\begin{algorithmic}[1]
		\STATE \textbf{Initialization:} Initialize the starting (or end) point $\qq(0)$, the maximum speed $v_{\text{max}}$, maximum acceleration $a_{\text{max}}$, positions of items $\ww_i,\forall i \in [K]$, boundary of obstacles $\mathbf{p}_{\text{min}}^i$ and $\mathbf{p}_{\text{max}}^i$;
		\STATE Determine the sequence with LKH by solving \eqref{10pro};
		\STATE Determine the flight trajectories between the items and the flight trajectory from the last item to the destination by executing Algorithm \ref{algo1};
		\STATE \textbf{Output:} The overall trajectory $\qq[1],\cdots,\qq[N]$.
	\end{algorithmic}
\end{algorithm}

\section{Numerical Results}
In this section, we present simulation results to evaluate the proposed LKH-DDPG algorithm for UAV trajectory design for logistics collection tasks in the LAE scenario. 

\subsection{Simulation Configuration}
The system parameters are set as follows. The UAV operates in a 3D space with the size of $1000 \times 1000 \times 1000~\text{m}^3$. The UAV starts from and returns to the workstation located at $(0,0,0)$. Its maximum flight speed and acceleration are  $v_{\text{max}}=50~\text{m/s}$ and $a_{\text{max}}=20~\text{m/s}^2$, respectively. The flight duration is discretized into time slots with a length of $\delta=1~\text{s}$. The collecting threshold $ \epsilon$ is set to $5$ m.  Obstacles are randomly generated within the 3D space as rectangular cuboids, and the items to be collected are randomly distributed in the 3D space. To encourage the UAV to efficiently collect items while avoiding collisions with obstacles, the reward function is designed as follows:
\(r_1=-\|\mathbf{q}[n]-\mathbf{w}_i\|\), \(r_2=100\), \(r_3=-\|\mathbf{q}[n]-\mathbf{q}[1]\|\), \(r_4=100\), \(r_5=-100\), and  \(r_6=-200\).

\subsection{Performance Benchmarks}
We consider two baselines that represent widely used approaches for UAV trajectory design in complex environments: 

The first baseline is {DDPG without LKH}, in which the UAV's trajectory is optimized solely by the DDPG algorithm~\cite{ShuyanXinTITS2024}, while the visiting sequence of items is determined randomly with no combinatorial optimization. 
In this setting, the reinforcement learning agent directly learns a continuous control policy that adapts the UAV's trajectory online based on the observed environment state, which includes its own position, velocity, relative distances to items, and the presence of obstacles. 
Since the visiting order of items is not optimized, the UAV may take inefficient routes that involve redundant detours or revisiting certain regions unnecessarily. 
This baseline is representative of a standard DRL strategy that focuses on local trajectory adaptation without exploiting global combinatorial structure in the collection order. 
By comparing with this baseline, we can quantify the performance gain of the LKH-based visiting sequence optimization in our framework.

The second baseline is an MPC scheme, which is a classical control-based approach widely used in robotics and autonomous systems for real-time trajectory planning~\cite{SavkinTIV2023}. 
MPC formulates the UAV trajectory planning problem as a finite-horizon optimal control problem, where the UAV's motion is modeled using discrete-time kinematics. Constraints, such as maximum velocity, maximum acceleration, and strict obstacle avoidance, are explicitly incorporated. 
In each decision step, MPC predicts the UAV's motion over a fixed prediction horizon towards the next target item, using the current state as the initial condition. 
It then solves a constrained optimization problem that minimizes a cost function combining multiple terms, including the travel distance to the next item, smoothness of the trajectory (via penalizing control effort), and proximity penalties for approaching obstacles. 
The optimization is subject to kinematic constraints (e.g., \(\|\mathbf{v}[n]\| \leq v_{\max},\) and \( \|\mathbf{a}[n]\| \leq a_{\max}\)) and geometric constraints ensuring the UAV remains outside all obstacle boundaries. 
Only the first control action from the optimized sequence is applied to the UAV, after which the system state is updated based on the executed action and environment feedback. 
This receding-horizon procedure is repeated until all items are collected and the UAV returns to the depot.

For the proposed LKH-DDPG algorithm, MPC can serve as a strong classical control-based baseline that is capable of generating dynamically feasible and smooth trajectories while explicitly accounting for physical constraints and obstacle avoidance. 
Since MPC is implemented with a fixed pre-defined visiting sequence of items (rather than optimizing it jointly with the trajectory), it may suffer from suboptimal global routing compared to the LKH-DDPG method. 
To this end, the LKH-DDPG framework has the following two strengths: (i) the ability to jointly optimize the global visiting order (via LKH) and local continuous trajectories (via DDPG), and (ii) the adaptability to highly non-convex feasible regions with dynamic obstacle-avoidance requirements.

\begin{figure}[htbp]
	\centering
	\includegraphics[width=\columnwidth]{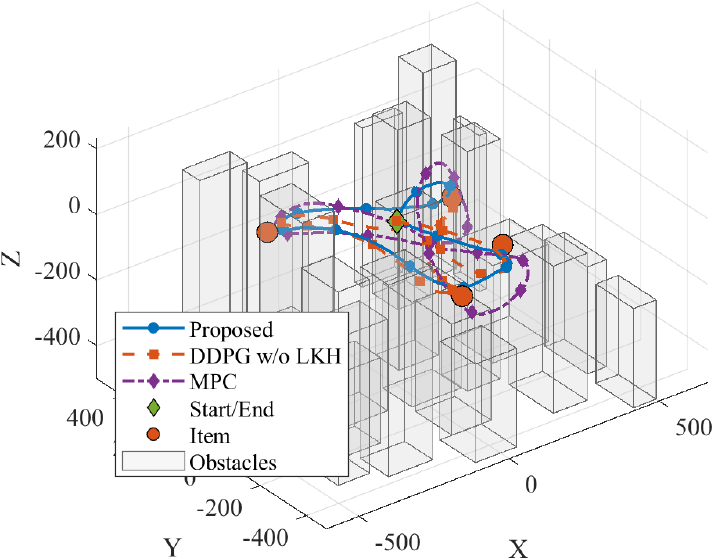}
	\caption{UAV trajectories in 3D view under the proposed LKH-DDPG, DDPG without LKH, and MPC schemes (scenario 1).}
	\label{fig:traj_2}
\end{figure}

\begin{figure}[htbp]
	\centering
	\includegraphics[width=\columnwidth]{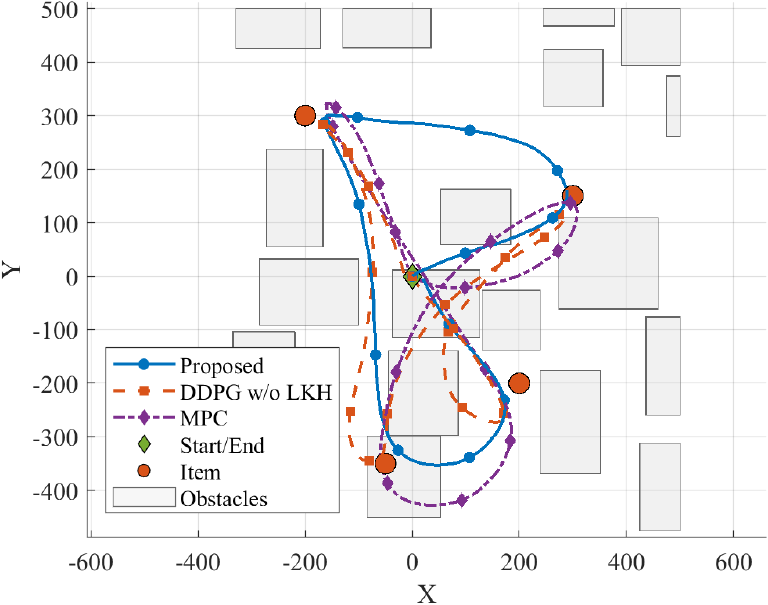}
	\caption{UAV trajectories in 2D top view under the proposed LKH-DDPG, DDPG without LKH, and MPC schemes (scenario 1).}
	\label{fig:traj_2_z}
\end{figure}

\subsection{Performance Evaluation}
Two scenarios, namely, \textit{Scenarios 1} and \textit{2}, are considered in our simulation:

In \textit{Scenario~1}, the arrangement of obstacles and item positions is shown in Figs.~\ref{fig:traj_2} and~\ref{fig:traj_2_z}, respectively, which illustrate the UAV's trajectories generated by the proposed LKH-DDPG, DDPG without LKH, and MPC schemes in both 3D and top views. The corresponding mission completion times are $71.8$~s for the proposed LKH-DDPG scheme, $87.4$~s for DDPG without LKH, and $105.6$~s for MPC. The  LKH-DDPG method shortens the mission time by approximately $17.9\%$, compared to DDPG without LKH, and by $32.0\%$, compared to MPC.

In Figs.~\ref{fig:traj_2} and \ref{fig:traj_2_z}, the baseline DDPG without LKH exhibits a non-optimal visiting order, resulting in redundant detours and a longer flight path and leading to a longer mission time. In contrast, the proposed LKH-DDPG jointly optimizes the global visiting sequence and the local continuous trajectory, producing smoother paths with fewer unnecessary turns and efficient obstacle avoidance, thus achieving the shortest mission time. The MPC baseline generates dynamically feasible and smooth trajectories that respect kinematic and obstacle-avoidance constraints; however, its global path is still suboptimal since it relies on a fixed, pre-defined visiting order and optimizes only the local motion toward the next target, leading to a longer mission time than LKH-DDPG and even longer than DDPG without LKH.

\begin{figure}[t]
	\centering
	\includegraphics[width=\columnwidth]{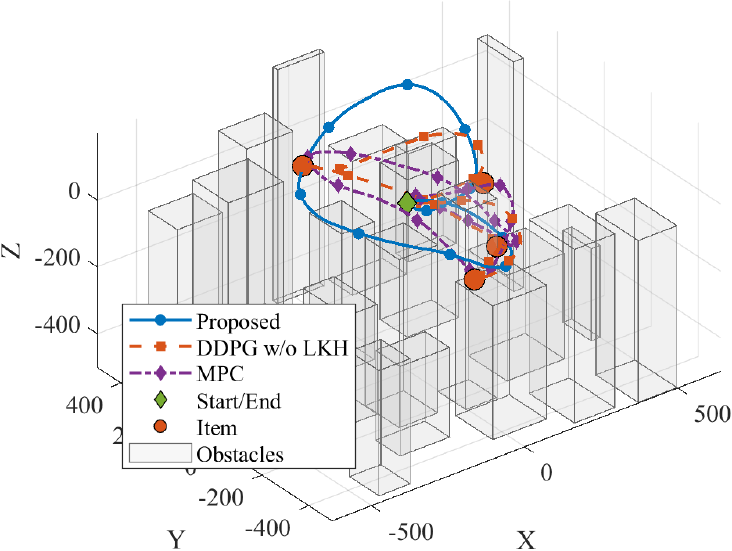}
	\caption{UAV trajectories in 3D view under the proposed LKH-DDPG, DDPG without LKH, and MPC schemes (scenario 2).}
	\label{fig:traj_3}
\end{figure}

\begin{figure}[htbp]
	\centering
	\includegraphics[width=\columnwidth]{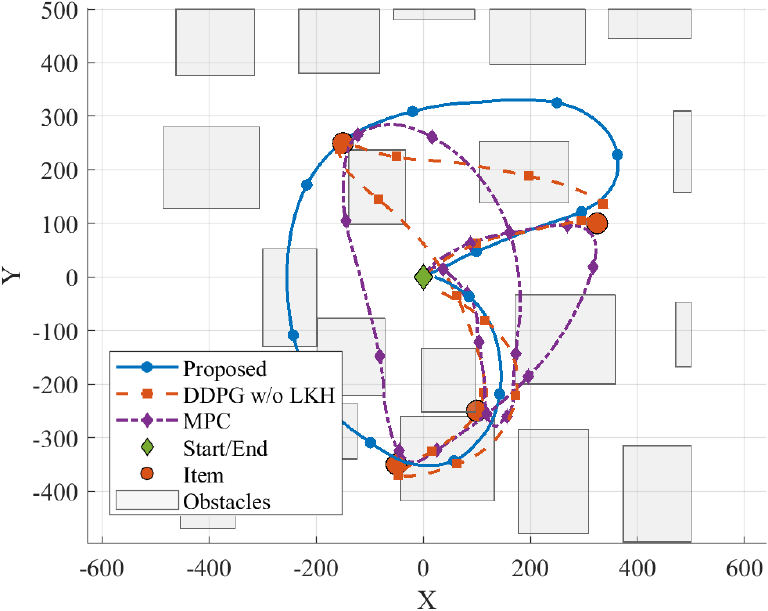}
	\caption{UAV trajectories in 2D top view under the proposed LKH-DDPG, DDPG without LKH, and MPC schemes (scenario 2).}
	\label{fig:traj_3_z}
\end{figure}

In \textit{Scenario~2}, the arrangement of obstacles and item positions is altered from those in Scenario~1 to create a different environment. The UAV's trajectories obtained under the proposed LKH-DDPG, the baseline DDPG without LKH, and MPC schemes are shown in Figs.~\ref{fig:traj_3} and~\ref{fig:traj_3_z} for the 3D and top views, respectively.

It is observed that the proposed LKH-DDPG scheme produces smooth and efficient paths that avoid obstacles with minimal detours, benefiting from the globally optimized visiting sequence provided by the LKH algorithm. In this scenario, the LKH-DDPG method achieves a mission completion time of $74.8$~s, which is $21.3\%$ shorter than DDPG without LKH ($95.0$~s) and $9.2\%$ shorter than MPC ($82.4$~s). In contrast, the DDPG without LKH lacks such global optimization, resulting in longer and less direct trajectories with several redundant turns and crossings visible from the top view. The MPC scheme achieves safe navigation by reactively avoiding obstacles; however, its short-horizon planning leads to conservative maneuvers and a longer travel distance than the proposed LKH-DDPG method.

These results are consistent with those from Scenario~1, confirming that integrating LKH-based sequence optimization with DDPG not only shortens the mission time but also enhances navigation efficiency, even in environments with significantly different obstacle and item layouts.

\begin{figure}[htbp]
	\centering
	\includegraphics[width=\columnwidth]{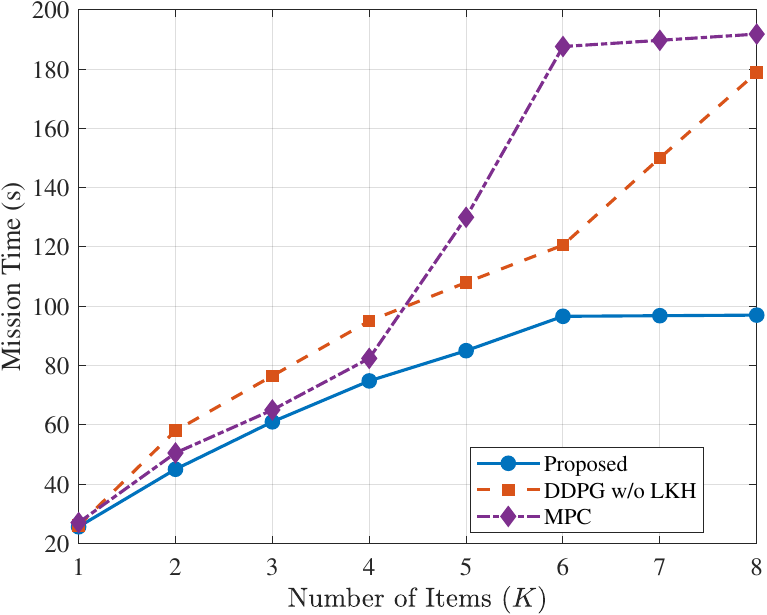}
	\caption{Mission completion time versus the number of items $K$ under the proposed LKH-DDPG, DDPG without LKH, and MPC schemes.}
	\label{fig:diff_item_time}
\end{figure}

Fig.~\ref{fig:diff_item_time} plots the mission completion time versus the number of items $K$ for the three considered schemes. 
The proposed LKH-DDPG consistently achieves the shortest time across all settings; DDPG without LKH increases steadily with $K$, while MPC is competitive for small $K$ but its mission completion time grows sharply when $K\ge5$.

When $K=1$,  all methods only need to depart from and return to the depot, hence the times are very close (explicitly 25.6~s, 26.0~s, and 27.0~s for the proposed LKH-DDPG, DDPG without LKH, and MPC, respectively). 
When $2\leq K\leq4$, the proposed method demonstrates clear gains (e.g.,  $74.8$~s vs.\ $95.0$~s for DDPG w/o LKH and $82.4$~s for MPC at $K=4$). 
The improvement comes from global sequence optimization by LKH that shortens the tour and from local DDPG smoothing that reduces unnecessary braking/turning near obstacles; MPC, solving a receding-horizon optimal control toward the next target with accurate kinematic and obstacle constraints, remains a strong classical baseline and thus outperforms the DRL-only baseline at small $K$.

As $K$ grows further, the difference between whether the global visiting order is optimized or not becomes dominant. 
For DDPG without LKH, the random visiting sequence induces self-crossings and backtracking, lengthening the path and causing repeated deceleration/acceleration near obstacles; the time grows from $58.0$~s ($K=2$) to $178.8$~s ($K=8$). 
For MPC, the finite-horizon planning toward only the next item and the fixed (predefined) visiting order lead to myopic detours when the environment is cluttered; once $K\ge5$, avoidance maneuvers and back-and-forth motions accumulate, producing a pronounced knee in the curve (from $82.4$~s for $K=4$ to $130.0\sim191.8$~s for $5\leq K \leq 8$). 
In contrast, the proposed LKH-DDPG couples a near-shortest global tour with locally learned, obstacle-aware continuous trajectories. The additional items beyond $K\!\approx\!5$ lie close to the already efficient loop and only incur marginal detours; correspondingly, the mission time plateaus around $96$–$97$~s for $6\leq K\leq 8$.

In summary, when the visiting order is not optimized (DDPG w/o LKH) or is fixed and planned myopically (MPC), the flight time grows rapidly with $K$ due to compounding detours and turn-induced slowdowns near obstacles. 
Jointly optimizing the global collection sequence (LKH) and the local obstacle-aware trajectory (DDPG) yields smoother paths with fewer turns and minimal backtracking, translating into substantial time reduction. For example, at $K=8$, the proposed method reduces the mission time by about $46\%$ compared to DDPG w/o LKH (97.0 vs.\ 178.8~s), and about $49\%$ compared to MPC (97.0 vs.\ 191.8~s). 
These results validate that global–local co-design is crucial for scalability in obstacle-rich low-altitude logistics scenarios.

\section{Conclusion}
This paper proposed an effective UAV trajectory planning framework for item collection in obstacle-rich LAE. 
We designed a novel LKH-DDPG algorithm that integrates the LKH algorithm for global visiting order optimization with DDPG for local continuous trajectory adaptation. 
The LKH component ensures a near-optimal global visiting order, while the DDPG component generates feasible and smooth flight paths that satisfy kinematic and obstacle-avoidance constraints. 
Simulations demonstrated that the proposed LKH-DDPG scheme consistently achieves shorter mission completion times and more efficient paths than control-theoretic and DRL baselines, particularly as the number of items increases.
The findings highlight the importance of combining global combinatorial optimization with local trajectory refinement for time-critical UAV missions in complex 3D environments. 
Future work will extend the LKH-DDPG to multi-UAV cooperative scenarios and real-world flight experiments.

\bibliographystyle{IEEEtran}
\bibliography{references}

\begin{thebibliography}{10}
\providecommand{\url}[1]{#1}
\csname url@samestyle\endcsname
\providecommand{\newblock}{\relax}
\providecommand{\bibinfo}[2]{#2}
\providecommand{\BIBentrySTDinterwordspacing}{\spaceskip=0pt\relax}
\providecommand{\BIBentryALTinterwordstretchfactor}{4}
\providecommand{\BIBentryALTinterwordspacing}{\spaceskip=\fontdimen2\font plus
\BIBentryALTinterwordstretchfactor\fontdimen3\font minus
  \fontdimen4\font\relax}
\providecommand{\BIBforeignlanguage}[2]{{%
\expandafter\ifx\csname l@#1\endcsname\relax
\typeout{** WARNING: IEEEtran.bst: No hyphenation pattern has been}%
\typeout{** loaded for the language `#1'. Using the pattern for}%
\typeout{** the default language instead.}%
\else
\language=\csname l@#1\endcsname
\fi
#2}}
\providecommand{\BIBdecl}{\relax}
\BIBdecl

\bibitem{watson2020maximizing}
T.~Watson, ``Maximizing the value of {America}'s newest resource, low-altitude
  airspace: An economic analysis of aerial trespass and drones,'' \emph{Ind.
  LJ}, vol.~95, p. 1399, 2020.

\bibitem{wu2024realization}
T.-L. Wu, C.-M. Kung, G.-T. Lin, K.-Y. Tsai, G.-Y. Meng, Y.-C. Huang, and W.-S.
  Cheng, ``The realization factors of {Taiwan}'s low altitude airspace economy
  and uam: Analysis and evaluation,'' in \emph{2024 IEEE International
  Conference on e-Business Engineering (ICEBE)}.\hskip 1em plus 0.5em minus
  0.4em\relax IEEE, 2024, pp. 1--4.

\bibitem{xiong2024evtol}
K.~Xiong, J.~Xie, and S.~Leng, ``{eVTOL} communication and trajectory
  optimization in low-altitude economy,'' in \emph{2024 IEEE/CIC International
  Conference on Communications in China (ICCC Workshops)}.\hskip 1em plus 0.5em
  minus 0.4em\relax IEEE, 2024, pp. 845--850.

\bibitem{yang2025intelligent}
H.~Yang, M.~Zheng, Z.~Shao, Y.~Jiang, and Z.~Xiong, ``Intelligent computation
  offloading and trajectory planning for {3D} target search in low-altitude
  economy scenarios,'' \emph{IEEE Wireless Commun. Lett.}, 2025.

\bibitem{tang2021systematic}
J.~Tang, S.~Lao, and Y.~Wan, ``Systematic review of collision-avoidance
  approaches for unmanned aerial vehicles,'' \emph{IEEE Syst. J.}, vol.~15,
  no.~3, pp. 3045--3057, 2021.

\bibitem{hsu2022rl-uav_comm}
Y.-H. Hsu and R.-H. Gau, ``Reinforcement learning-based collision avoidance and
  optimal trajectory planning in {UAV} communication networks,'' \emph{IEEE
  Trans. Mob. Comput.}, vol.~21, no.~1, pp. 306--320, 2020.

\bibitem{wang2021learning_collision}
X.~Wang and M.~C. Gursoy, ``Learning-based {UAV} trajectory optimization with
  collision avoidance and connectivity constraints,'' \emph{IEEE Trans.
  Wireless Commun.}, vol.~21, no.~6, pp. 4350--4363, 2021.

\bibitem{gong2022bor_opt}
S.~Gong, M.~Wang, B.~Gu, W.~Zhang, D.~T. Hoang, and D.~Niyato, ``Bayesian
  optimization enhanced deep reinforcement learning for trajectory planning and
  network formation in multi-{UAV} networks,'' \emph{IEEE Trans. Veh.
  Technol.}, vol.~72, no.~8, pp. 10\,933--10\,948, 2023.

\bibitem{sha2025rrt_sa_cost}
X.-H. Shao and Z.~Chen, ``Path planning algorithm for {UAV}s in dynamic
  adversarial environments based on {RRT-SAC},'' \emph{IEEE J. Sel. Areas
  Commun.}, vol.~43, no.~6, pp. 1300--1312, 2025.

\bibitem{xu2021dpm_pc}
Z.~Xu, D.~Deng, Y.~Dong, and K.~Shimada, ``{DPMPC}-planner: A real-time {UAV}
  trajectory planning framework for complex static environments with dynamic
  obstacles,'' \emph{IEEE Trans. Cybern.}, pp. 250--256, 2022.

\bibitem{roghair2021vision_rl}
J.~Roghair, A.~Niaraki, K.~Ko, and A.~Jannesari, ``A vision based deep
  reinforcement learning algorithm for {UAV} obstacle avoidance,'' in
  \emph{Proc. SAI Intell. Syst. Conf.}\hskip 1em plus 0.5em minus 0.4em\relax
  Springer, 2021, pp. 115--128.

\bibitem{fahlstrom2022introduction}
P.~G. Fahlstrom, T.~J. Gleason, and M.~H. Sadraey, \emph{Introduction to {UAV}
  systems}.\hskip 1em plus 0.5em minus 0.4em\relax John Wiley \& Sons, 2022.

\bibitem{zhai2025integrated}
Z.~Zhai, W.~Ni, X.~Wang, D.~Niyato, and E.~Hossain, ``Integrated sensing and
  communication with {UAV} swarms via decentralized consensus admm,''
  \emph{arXiv preprint arXiv:2511.03283}, 2025.

\bibitem{zhai2024uav}
Z.~Zhai, X.~Yuan, and X.~Wang, ``{UAV}-enabled over-the-air asynchronous
  federated learning,'' in \emph{Proc. Int. Conf. Wireless Commun. Signal
  Process. (WCSP)}.\hskip 1em plus 0.5em minus 0.4em\relax IEEE, 2024, pp.
  740--745.

\bibitem{zhang2023threat_model}
J.~Li, Y.~Xiong, and J.~She, ``{UAV} path planning for target coverage task in
  dynamic environment,'' \emph{IEEE Internet Things J.}, vol.~10, no.~20, pp.
  17\,734--17\,745, 2023.

\bibitem{tu2023q_sarsa}
G.-T. Tu and J.-G. Juang, ``{UAV} path planning and obstacle avoidance based on
  reinforcement learning in {3D} environments,'' in \emph{Actuators}, vol.~12,
  no.~2.\hskip 1em plus 0.5em minus 0.4em\relax MDPI, 2023, p.~57.

\bibitem{shaoxiao2025airspace_comm}
H.~Oh, H.-S. Shin, S.~Kim, and W.-H. Chen, ``Communication-aware trajectory
  planning for unmanned aerial vehicles in urban environments,'' \emph{J. Guid.
  Control Dyn.}, vol.~41, no.~10, pp. 2271--2282, 2018.

\bibitem{9169676}
Y.~Liu, S.~Xie, and Y.~Zhang, ``Cooperative offloading and resource management
  for {UAV}-enabled mobile edge computing in power iot system,'' \emph{IEEE
  Trans. Veh. Technol.}, vol.~69, no.~10, pp. 12\,229--12\,239, 2020.

\bibitem{wu2018joint}
Q.~Wu, Y.~Zeng, and R.~Zhang, ``Joint trajectory and communication design for
  multi-{UAV} enabled wireless networks,'' \emph{IEEE Trans. Wireless Commun.},
  vol.~17, no.~3, pp. 2109--2121, 2018.

\bibitem{zhai2022energy}
Z.~Zhai, X.~Dai, B.~Duo, X.~Wang, and X.~Yuan, ``Energy-efficient {UAV}-mounted
  ris assisted mobile edge computing,'' \emph{IEEE Wireless Commun. Lett.},
  vol.~11, no.~12, pp. 2507--2511, 2022.

\bibitem{7888557}
Y.~Zeng and R.~Zhang, ``Energy-efficient {UAV} communication with trajectory
  optimization,'' \emph{IEEE Trans. Wireless Commun.}, vol.~16, no.~6, pp.
  3747--3760, 2017.

\bibitem{10818523}
Z.~Zhai, X.~Yuan, X.~Wang, and H.~Yang, ``{UAV}-enabled asynchronous federated
  learning,'' \emph{IEEE Trans. Wireless Commun.}, vol.~24, no.~3, pp.
  2358--2372, 2025.

\bibitem{sun2021joint}
C.~Sun, W.~Ni, and X.~Wang, ``Joint computation offloading and trajectory
  planning for uav-assisted edge computing,'' \emph{IEEE Trans. Wireless
  Commun.}, vol.~20, no.~8, pp. 5343--5358, 2021.

\bibitem{xiong2022three}
X.~Xiong, C.~Sun, W.~Ni, and X.~Wang, ``Three-dimensional trajectory design for
  unmanned aerial vehicle-based secure and energy-efficient data collection,''
  \emph{IEEE Transactions on Vehicular Technology}, vol.~72, no.~1, pp.
  664--678, 2022.

\bibitem{banerjee2021uav_risk}
P.~Banerjee, G.~Gorospe, and E.~Ancel, ``{3-D} representation of {UAV}-obstacle
  collision risk under off-nominal conditions,'' in \emph{Proc. IEEE Aerosp.
  Conf.}, 2021, pp. 1--11.

\bibitem{nzhang2021autonomous_collision}
J.~L. Sanchez-Lopez, M.~Wang, M.~A. Olivares-Mendez, M.~Molina, and H.~Voos,
  ``A real-time {3D} path planning solution for collision-free navigation of
  multirotor aerial robots in dynamic environments,'' \emph{J. Intell. Robot.
  Syst.}, vol.~93, no.~1, pp. 33--53, 2019.

\bibitem{10480601}
Y.~Gao, X.~Yuan, D.~Yang, Y.~Hu, Y.~Cao, and A.~Schmeink, ``{UAV}-assisted
  {MEC} system with mobile ground terminals: {DRL}-based joint terminal
  scheduling and {UAV} {3D} trajectory design,'' \emph{IEEE Trans. Veh.
  Technol.}, vol.~73, no.~7, pp. 10\,164--10\,180, 2024.

\bibitem{ma2023flightcorridor}
Z.~Ma, Z.~Wang, A.~Ma, Y.~Liu, and Y.~Niu, ``A low-altitude obstacle avoidance
  method for {UAV}s based on polyhedral flight corridor,'' \emph{IEEE Trans.
  Aerosp. Electron. Syst.}, vol.~59, no.~3, pp. 1600--1610, 2023.

\bibitem{chan2025nearoptimal}
Y.~Y. Chan, K.~K.~H. Ng, T.~Wang, K.~K. Hon, and C.-H. Liu, ``Near time-optimal
  trajectory optimisation for drones in last-mile delivery using spatial
  reformulation approach,'' \emph{Transp. Res. Part C Emerg. Technol.}, vol.
  171, p. 104986, 2025.

\bibitem{chao2025airspace}
D.~Chao, Y.~Zhang, Z.~Jia, Y.~Liao, L.~Zhang, and Q.~Wu, ``Three-dimension
  collision-free trajectory planning of {UAV}s based on {ADS}-{B} information
  in low-altitude urban airspace,'' \emph{Chin. J. Aeronaut.}, vol.~38, no.~2,
  p. 103170, 2025.

\bibitem{hu2023droneswarm}
J.~Hu, L.~Fan, Y.~Lei, Z.~Xu, W.~Fu, and G.~Xu, ``Reinforcement learning-based
  low-altitude path planning for {UAS} swarm in diverse threat environments,''
  \emph{IEEE Internet Things J.}, vol.~10, no.~5, pp. 4200--4213, 2023.

\bibitem{lyu2018uav_jsac}
J.~Lyu, Y.~Zeng, R.~Zhang, and T.~J. Lim, ``Placement optimization of
  {UAV}-mounted mobile base stations,'' \emph{IEEE Commun. Lett.}, vol.~21,
  no.~3, pp. 604--607, 2017.

\bibitem{zhang2019sec_twc}
G.~Zhang, Q.~Wu, M.~Cui, and R.~Zhang, ``Securing {UAV} communications via
  joint trajectory and power control,'' \emph{IEEE Trans. Wireless Commun.},
  vol.~18, no.~2, pp. 1376--1389, 2019.

\bibitem{helsgaun2015solving}
K.~Helsgaun, ``Solving the equality generalized traveling salesman problem
  using the {Lin}--{Kernighan}--{Helsgaun} algorithm,'' \emph{Math. Program.
  Comput.}, vol.~7, pp. 269--287, 2015.

\bibitem{lillicrap2015continuous}
T.~P. Lillicrap, J.~J. Hunt, A.~Pritzel, N.~Heess, T.~Erez, Y.~Tassa,
  D.~Silver, and D.~Wierstra, ``Continuous control with deep reinforcement
  learning,'' \emph{arXiv preprint arXiv:1509.02971}, 2015.

\bibitem{boyd2004convex}
S.~P. Boyd and L.~Vandenberghe, \emph{Convex Optimization}.\hskip 1em plus
  0.5em minus 0.4em\relax Cambridge university press, 2004.

\bibitem{liu2019stochastic}
A.~Liu, V.~K. Lau, and B.~Kananian, ``Stochastic successive convex
  approximation for non-convex constrained stochastic optimization,''
  \emph{IEEE Trans. Signal Process.}, vol.~67, no.~16, pp. 4189--4203, 2019.

\bibitem{marini2022reinforcement}
R.~Marini, L.~Spampinato, S.~Mignardi, R.~Verdone, and C.~Buratti,
  ``Reinforcement learning-based trajectory planning for {UAV}-aided vehicular
  communications,'' in \emph{Proc. Eur. Signal Process. Conf. (EUSIPCO)}.\hskip
  1em plus 0.5em minus 0.4em\relax IEEE, 2022, pp. 967--971.

\bibitem{ning2023multi}
Z.~Ning, Y.~Yang, X.~Wang, Q.~Song, L.~Guo, and A.~Jamalipour, ``Multi-agent
  deep reinforcement learning based {UAV} trajectory optimization for
  differentiated services,'' \emph{IEEE Trans. Mob. Comput.}, vol.~23, no.~5,
  pp. 5818--5834, 2023.

\bibitem{gong2023bayesian}
S.~Gong, M.~Wang, B.~Gu, W.~Zhang, D.~T. Hoang, and D.~Niyato, ``Bayesian
  optimization enhanced deep reinforcement learning for trajectory planning and
  network formation in multi-{UAV} networks,'' \emph{IEEE Trans. Veh.
  Technol.}, vol.~72, no.~8, pp. 10\,933--10\,948, 2023.

\bibitem{roghair2021vision}
J.~Roghair, A.~Niaraki, K.~Ko, and A.~Jannesari, ``A vision based deep
  reinforcement learning algorithm for {UAV} obstacle avoidance,'' in
  \emph{Proc. SAI Intell. Syst. Conf.}\hskip 1em plus 0.5em minus 0.4em\relax
  Springer, 2021, pp. 115--128.

\bibitem{yan2020towards}
C.~Yan, X.~Xiang, and C.~Wang, ``Towards real-time path planning through deep
  reinforcement learning for a {UAV} in dynamic environments,'' \emph{J.
  Intell. Robot. Syst.}, vol.~98, no.~2, pp. 297--309, 2020.

\bibitem{hoffman2013traveling}
K.~L. Hoffman, M.~Padberg, G.~Rinaldi \emph{et~al.}, ``Traveling salesman
  problem,'' \emph{Encycl. Oper. Res. Manag. Sci.}, vol.~1, pp. 1573--1578,
  2013.

\bibitem{sawik2016note}
T.~Sawik, ``A note on the {Miller}-{Tucker}-{Zemlin} model for the asymmetric
  traveling salesman problem,'' \emph{Bull. Polish Acad. Sci. Tech. Sci.}, pp.
  517--520, 2016.

\bibitem{ShuyanXinTITS2024}
S.~Hu, X.~Yuan, W.~Ni, X.~Wang, and A.~Jamalipour, ``Visual-based moving target
  tracking with solar-powered fixed-wing {UAV:} {A} new learning-based
  approach,'' \emph{{IEEE} Trans. Intell. Transp. Syst.}, vol.~25, no.~8, pp.
  9115--9129, 2024.

\bibitem{SavkinTIV2023}
M.~Eskandari, H.~Huang, A.~V. Savkin, and W.~Ni, ``Model predictive
  control-based {3D} navigation of a {RIS}-equipped {UAV} for los wireless
  communication with a ground intelligent vehicle,'' \emph{{IEEE} Trans.
  Intell. Veh.}, vol.~8, no.~3, pp. 2371--2384, 2023.

\end{thebibliography}

\end{document}